\documentclass[superscriptaddress,11pt]{revtex4-1}
\usepackage[utf8]{inputenc}
\usepackage[english]{babel}
\usepackage{graphicx}
\usepackage{amsmath,amssymb}
\usepackage{hyperref}

\newcommand{\nagoya}{Graduate School of Informatics, Nagoya University, Japan}
\newcommand{\jst}{JST, PRESTO, Japan}
\newcommand{\cnets}{Center for Complex Networks and Systems Research, Indiana University Bloomington, USA}
\newcommand{\sice}{Luddy School of Informatics, Computing, and Engineering, Indiana University, Bloomington, USA}
\newcommand{\iuni}{Network Science Institute, Indiana University, USA}
\newcommand{\umsi}{School of Information, University of Michigan, USA}
\newcommand{\usf}{Department of Computer Science and Engineering, University of South Florida, USA}

\begin{document}

\title{Social Influence and Unfollowing Accelerate the Emergence of Echo Chambers}

\author{Kazutoshi Sasahara\footnote{Correspondence should be addressed to K.S. (sasahara@nagoya-u.jp)}}
\affiliation{\nagoya{}}
\affiliation{\jst{}}
\affiliation{\cnets{}}
\author{Wen Chen}
\affiliation{\cnets{}}
\affiliation{\sice{}}
\author{Hao Peng}
\affiliation{\cnets{}}
\affiliation{\umsi{}}
\author{Giovanni Luca Ciampaglia}
\affiliation{\iuni{}}
\affiliation{\usf{}}
\author{Alessandro Flammini}
\author{Filippo Menczer}
\affiliation{\cnets{}}
\affiliation{\sice{}}
\affiliation{\iuni{}}

\begin{abstract}
While social media make it easy to connect with and access information from anyone, they also facilitate basic influence and unfriending mechanisms that may lead to segregated and polarized clusters known as ``echo chambers.'' Here we study the conditions in which such echo chambers emerge by introducing a simple model of information sharing in online social networks with the two ingredients of influence and unfriending. Users can change both their opinions and social connections based on the information to which they are exposed through sharing. The model dynamics show that even with minimal amounts of influence and unfriending, the social network rapidly devolves into segregated, homogeneous communities. These predictions are consistent with empirical data from Twitter. Although our findings suggest that echo chambers are somewhat inevitable given the mechanisms at play in online social media, they also provide insights into possible mitigation strategies.
\end{abstract}

\maketitle

\section*{Introduction}

The rise of social media has led to unprecedented changes in the scale and speed at which people share information. Social media feeds are key tools for accessing high volumes of news, opinions, and public information. However, just by fostering such a proliferation of information to which people are exposed, social media may interfere with cognitive selection biases, amplifying undesirable phenomena such as extremism and the spread of misinformation~\cite{HillsProliferation18}. Further, they may introduce new biases in the way people consume information and form beliefs, which are not well understood yet.

Theories about group decision-making and problem-solving suggest that aggregating knowledge, insights, or expertise from a \emph{diverse} group of people is an effective strategy to solve complex problems, a notion called \emph{collective intelligence}~\cite{Bonabeau2009,page2008difference}. While the Web and social media have often been hailed as striking examples of this principle in action~\cite{surowiecki2005wisdom,benkler2006wealth}, some of the assumptions upon which these systems are predicated may harm the very diversity that makes them precious sources of collective intelligence~\cite{salganik2006experimental}. Social media mechanisms, in particular, tend to use popularity signals as proxies of quality or personal preference, despite evidence that rewarding engagement may come at the expense of viewpoint diversity and quality~\cite{Nematzadeh2017popularity-bias}. 
Worse, the community structure of information flow networks can distort decisions and increase vulnerability to malicious actors such as social bots~\cite{gerrymandering2019}.

\begin{figure}[t]
\includegraphics[width=\textwidth]{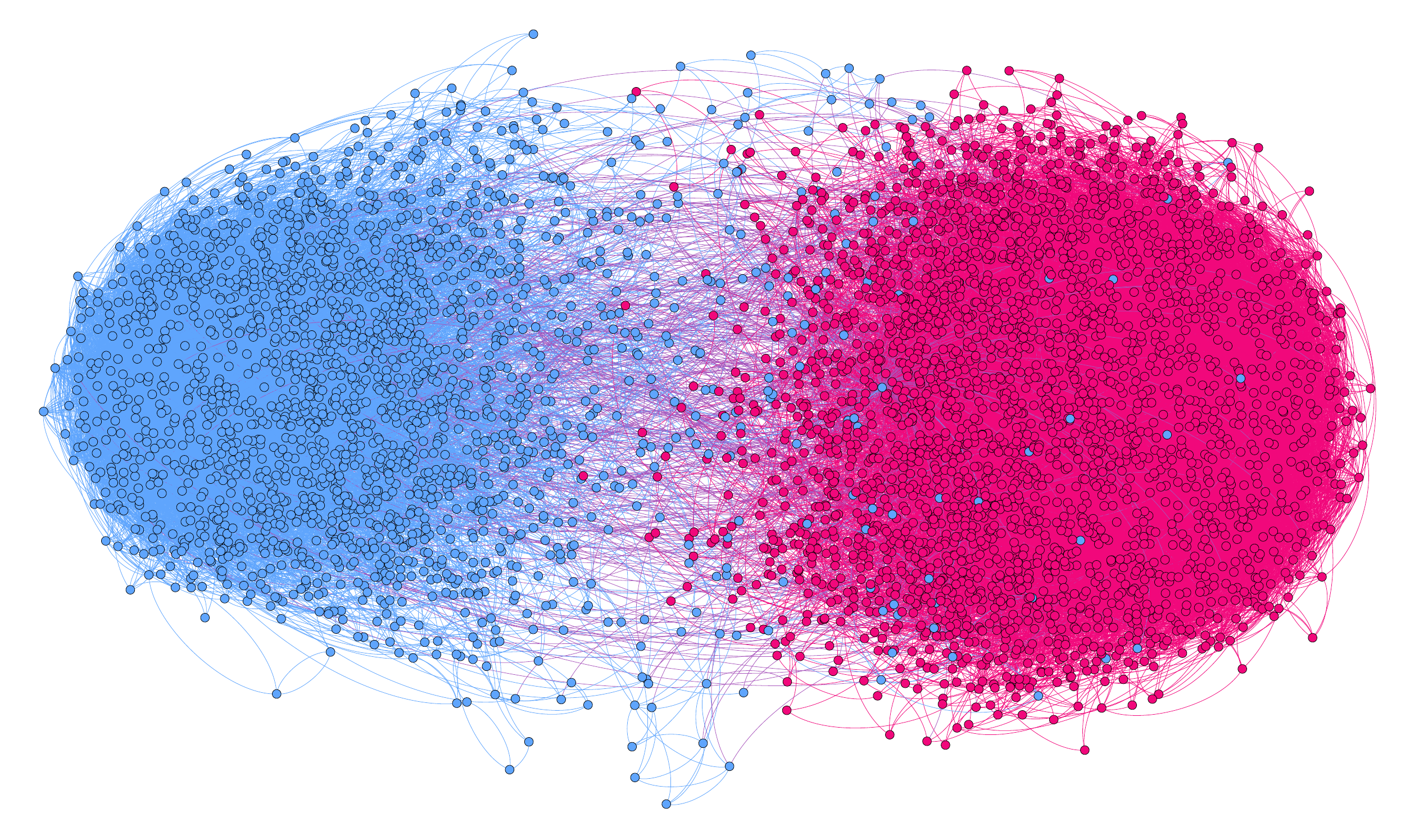}
\caption{Example of a polarized and segregated network on Twitter. The network visualizes retweets of political hashtags from the 2010 US midterm elections. 
The nodes represent Twitter users and there is a directed edge from node $i$ to node $j$ if user $j$ retweeted user $i$. Colors represent political preference: red for conservatives and blue for progressives~\cite{Conover2010predicting}. For illustration purposes, only the nodes in the $k=3$ core are visualized. See Methods for more details.}
\label{fig:echocham}
\end{figure}

There is increasing empirical evidence of these phenomena: polarization is observed in social media conversations~\cite{Conover2011,Conover2012,Bright2016} and low diversity is found in online news consumption~\cite{Bakshy2015,DelVicario2016a,flaxman2016filter,Schmidt2017}. These observations have in common two features: \emph{network segregation} (the splitting of the social network in two or more disconnected or poorly connected groups) and \emph{opinion polarization} (the high homogeneity of opinions within such groups). Fig.~\ref{fig:echocham} shows what an information diffusion network with those two features looks like. 
Human factors such as \emph{homophily}~\cite{marsden_homogeneity_1988,McPherson2001,centola_experimental_2011} (the tendency to form ties with similar people) and \emph{social influence}~\cite{friedkin2006structural} (the tendency of becoming more similar to somebody as a result of social interaction) are often thought to drive the emergence of polarization and segregation. 

Some of the consequences of socio-cognitive biases have been explored in social dynamics models~\cite{Castellano2009} and in the social psychology literature~\cite{HillsProliferation18}. Yet, the interplay between these and additional biases introduced by social media mechanisms is not clear. The algorithms at the heart of social media make a number of assumptions to deliver their recommendations. For example, news feed algorithms favor stories with which users are more likely to engage in the future, based on past engagement~\cite{Backstrom2016}. Friendship recommendation engines suggest new ties based on common interests, beliefs, and friends, often resulting in the closure of open triads~\cite{Adamic2003,Backstrom2006,Leskovec2008,weng2013role}. And finally, social media empower users to dissolve ties that, although not by design, often tend to be the ones connecting them with those with whom they disagree~\cite{Sibona2011}. Indeed, estimates suggest that tie decay in social networks is a relatively common occurrence~\cite{raeder2011predictors,10.1145/1978942.1979104}. The Appendix provides further empirical evidence of the dissolution of social connections on Twitter.

By curating the information to which users are exposed and by facilitating their management of social ties, social media platforms may enhance homophily and \emph{confirmation bias}, the tendency to pay attention only to information that aligns with prior beliefs~\cite{Nickerson1998}. This would have the net effect of leading social media users to connect preferentially with like-minded individuals, which would then result in \emph{selective exposure} to only that information which conforms to their pre-existing beliefs, as opposed to more diverse points of view~\cite{Sears1967}. Ultimately, these dynamics would drive users of social media toward polarization and segregation~\cite{sunstein2017republic}, more so than users of legacy media like TV, radio, or newspapers, where social sharing and link management mechanisms are not at play~\cite{Gentzkow2011}. 

The risk that online communication networks could splinter into different `tribes' was recognized at the dawn of cyberspace, and given the name of \emph{cyber-balkanization}~\cite{katz1998struggle,sunstein2002republiccom} as an analogy to the well-known phenomena of cultural, racial, and ethnic segregation~\cite{schelling1971dynamic}. With the rise of modern personalization technologies, there has been renewed interest in understanding whether algorithmic bias is accelerating the fragmentation of society. The terms \emph{filter bubble}~\cite{pariser2011filter} and \emph{echo chamber}~\cite{Jamieson2008} have been coined to refer to two different algorithmic pathways to opinion fragmentation, both related to the way algorithms filter and rank information. The first refers to search engines~\cite{fortunato2006topical}, the second to social media feeds~\cite{nikolov2015measuring,Nikolov2018biases}. 

The literature about the effects of technology on echo chambers presents a complicated picture, with inconsistent and somewhat contradictory evidence~\cite{guess2018avoiding}. This is not entirely surprising when one considers the enormous heterogeneity in news sources, social media characteristics and usage, and human information-consumption behaviors. 
Lab experiments show that people consistently exhibit a preference for congenial information over uncongenial information, especially in the domain of politics~\cite{hart2009feeling}. 
Surveys find mixed evidence of partisan selectivity~\cite{stroud2010polarization,dubois2018echo}.
On social media, opinion-reinforcing information promotes news story exposure~\cite{Garrett2009echo}. 
Analysis of online consumption of political news by Facebook users shows that exposure to cross-cutting ideological information is reduced by comparable amounts when one considers platform algorithms and individual click choices~\cite{Bakshy2015}. 
Behavioral data on online consumption of political news suggest that selective exposure is mostly concentrated among a minority of the population --- those who drive most of the traffic to partisan sources~\cite{shore2016network,Guess-diet}. 
Even if polarized media consumption may not be the norm for most people, it is common among the most politically active,
knowledgeable, engaged, and influential individuals~\cite{garimella2018political}. This can favor the spread of misinformation from partisan media and increase animosity within the population. It is therefore important to understand how specific social media mechanisms may facilitate the formation of ideological echo chambers. 

Here, we study the emergence of joint opinion polarization and network segregation in online social media, specifically focusing on the interplay between the mechanisms of influence and unfriending. One novel contribution of our approach is to model how these mechanisms are mediated by the basic activity of information sharing in social media. Furthermore, we explore how biases in recommendation algorithms may exacerbate the dynamics of echo chambers. Although our model is idealized, it captures some key features of social media sharing --- limited attention, social influence, and social tie curation. In particular we investigate the role of selective unfollowing, which has not been studied in the literature. Through a series of simulations that compare different scenarios by exploring two key parameters, we find conditions under which opinion polarization and network segregation coevolve. This provides a generative mechanism to interpret the formation of echo chambers in social media. We also check the consistency of our model against empirical data from Twitter. This allows us to test the micro-level assumptions underpinning our model as well as its macro-level predictions~\cite{flache2017models}.

\section*{Model}

Let us introduce a model of opinion dynamics in an evolving social network. We incorporate various ingredients from models in the literature: information diffusion via social sharing~\cite{Weng2012}, opinion influence based on bounded confidence~\cite{Deffuant2000}, and selective rewiring of social ties~\cite{Holme2006}. 

The model begins with a random directed graph with $N$ nodes and $E$ directed edges, representing an online social network over which messages spread. Nodes represent social media users and edges represent follower ties, as in Twitter and Instagram. In the initial step, each user's \emph{opinion} ($o$) is randomly assigned a value in the interval $[-1, +1]$. Each user has a \emph{screen} that shows the most recent $l$ messages posted (or reposted) by friends being followed. A message conveys the identity and opinion value $m$ of the user who originated the post, together with the information about who reposted it. Users can see this information. A message's opinion $m$ is \emph{concordant} with an opinion $o$ if they are within a \emph{bounded confidence} distance $\epsilon$ (that is, when $|o - m| < \epsilon$). It is discordant otherwise. In addition, each user can unfollow a friend by rewiring the connection, thus following someone else in their place. These mechanisms allow us to capture two common ingredients of social media platforms: the possibility to share information with one's friends, and the possibility to form a new connections. 

At every time step $t$, user $i$ is selected at random, and sees messages on the screen that are posted or reposted by friends. The opinion of user $i$ then changes based on the concordant messages on the screen: 
\begin{equation} 
o_i(t+1) = o_i(t) + \mu \frac{\sum_{j=1}^{l}I_{\epsilon}(o_i(t), m_j)(m_j -
o_i(t))}{\sum_{j=1}^{l}I_{\epsilon}(o_i(t), m_j)}, 
\label{eq:learning}
\end{equation}
where $\mu$ is an influence strength parameter, the sum runs over the messages in $i$'s screen, and $I_{\epsilon}$ is an indicator function for concordant opinions based on the confidence bound $\epsilon$:
\begin{equation}
   I_{\epsilon}(o,m) = \left\{
    \begin{array}{ll}
      1 & \text{if } |o - m| < \epsilon \\
      0 & \text{otherwise.}
    \end{array}
  \right.
\end{equation}
Equation~\ref{eq:learning} provides a simple mechanism for modeling social influence based on an individual's tendency to favor information that is similar to their pre-existing opinions ($o_i \pm \epsilon$), such as the confirmation bias and selective exposure mentioned earlier. This is referred to as \emph{bounded social influence} and its breadth and strength are controlled by the parameters $\epsilon$ and $\mu$. Larger $\epsilon$ indicates broader-minded users, and larger $\mu$ indicates stronger social influence.

Two more actions may be taken by user $i$ at each time step $t$. First, with probability $p$, the user reposts a concordant message from the screen, if any are available; otherwise, with probability $1-p$, they post a new message reflecting their own opinion. Second, with probability $q$, the user selects a random discordant message from the screen, if one exists, and \emph{unfollows} the friend who (re)posted the message, following a new friend in their place. We explore three different \emph{rewiring strategies} for selecting the new friend: 
\begin{itemize}
    \item{\textbf{random:}} a user is selected at random among all nodes in the network that are not already friends of $i$'s;
    \item{\textbf{repost:}} a user is selected at random among the originators of reposts, if any are on $i$'s screen; 
    \item{\textbf{recommendation:}} a user is selected at random among non-friends who recently posted concordant messages.
\end{itemize}
Note that the size, density, and out-degree sequence of the network stay the same throughout each simulation, while the in-degree distribution can change over time.

As we mentioned earlier, our model incorporates various elements that have been explored in the literature. Both opinion dynamics and the rewiring of social ties (unfriending) are notably present in the model proposed by Holme and Newman~\cite{Holme2006}, which first explored the roles of the two mechanisms. There are however a few key differences between the model presented here and previous models. One is that in our model, opinions take continuous values and unfriending is based on bounded confidence. More importantly, when links are rewired, they do not necessarily select nodes with concordant opinion (this is only one of the three rewiring strategies we consider); rather, the targets of the selection are the links to be broken --- those outside the opinion confidence bound. Finally, our model aims to capture the crucial features of information diffusion in social sharing platforms, where influence may take place indirectly. Consider for example the following scenario: user A posts a piece of information that reflects A's opinion; user B reshares the message to their followers, which include user C. Now user C may be influenced by A's post, even though A and C are \emph{not} directly connected, and irrespective of whether C's opinion was concordant with or influenced by B's opinion. This indirect influence mechanism is asymmetric: the opinion of the consumer of a post changes, while the opinion of the originator of the post does not. The average opinion is therefore not conserved, unlike in the model proposed by Deffuant~\cite{Deffuant2000}.

\begin{figure}
    \centerline{\includegraphics[width=1.0\textwidth]{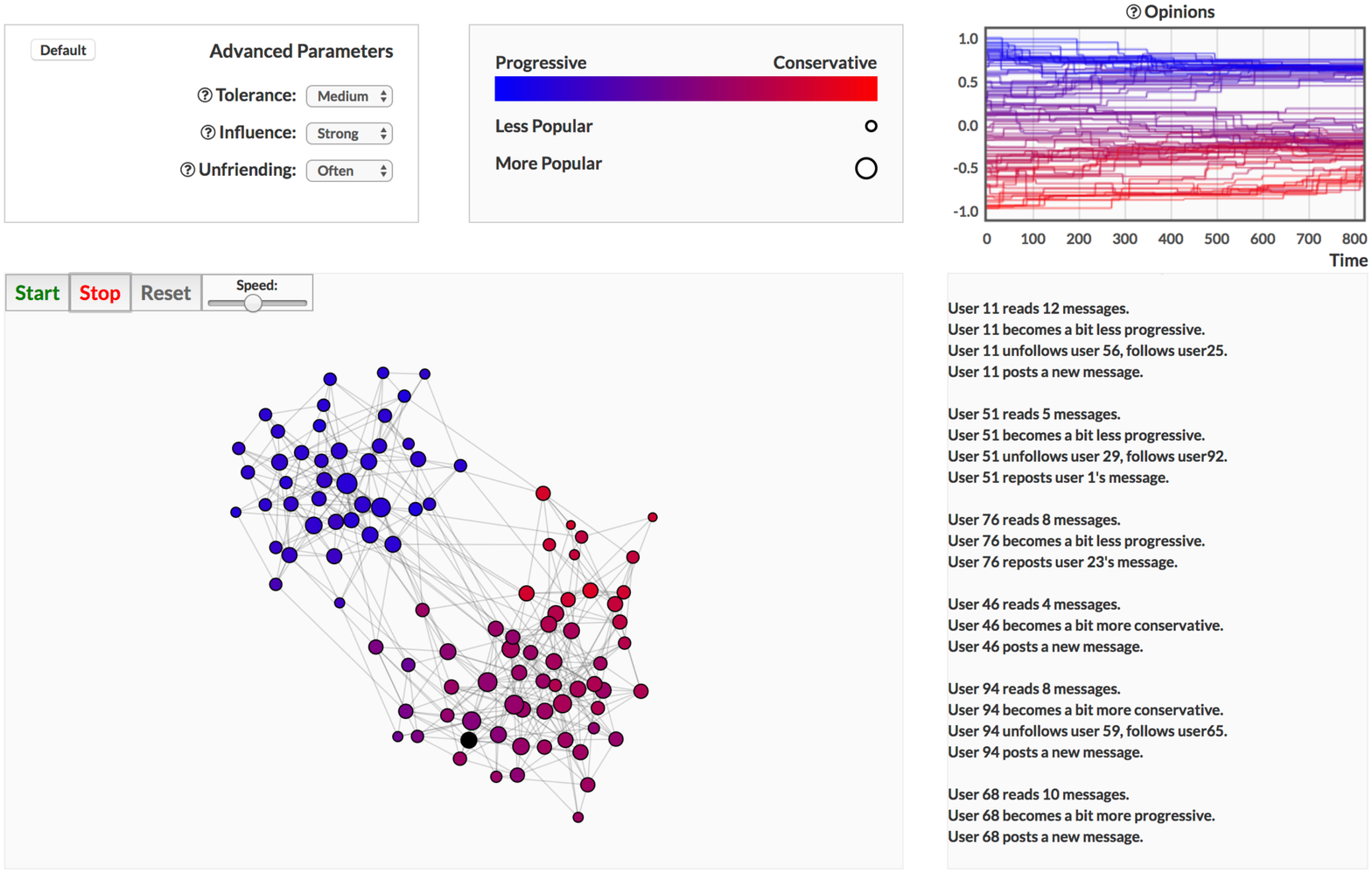}}
    \caption{Screenshot of the echo-chamber model demo.}
    \label{fig:demo}
\end{figure}

The code to simulate our model and reproduce our findings is available online at \url{github.com/soramame0518/echo_chamber_model}.

To facilitate the exploration of our model, we developed an interactive demo allowing one to run Web-based simulations with different realizations of the model parameters. The demo makes certain simplifications to be accessible to a broad audience: it is based on an undirected network, nodes can see all messages from their neighbors, and unfriending only occurs by random rewiring. Fig.~\ref{fig:demo} provides a screenshot of the demo, which is available online at \url{osome.iu.edu/demos/echo/}.

\section*{Results}

\subsection*{Emergence of Echo Chambers}

\begin{figure}[t]
\centerline{\includegraphics[width=1.0\textwidth]{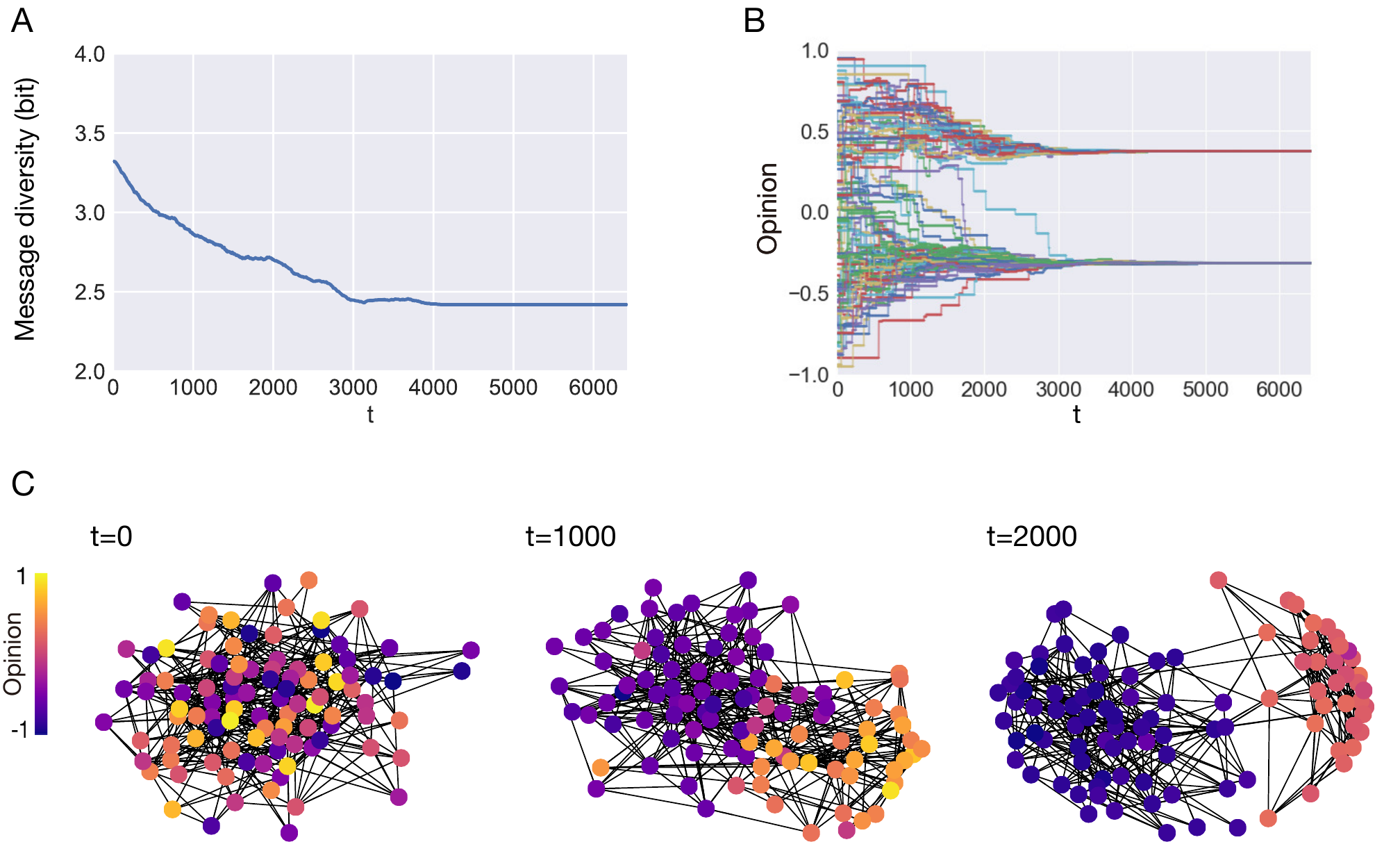}}
\caption{Coevolution of opinion polarization and network segregation. (A)~Average diversity of messages on the screen, measured using Shannon entropy with the opinion range divided into 10 bins. (B)~Temporal changes in population opinions. (C)~Temporal changes in the social network structure. We use parameters $N=100$, $E=400$, $\epsilon=0.4$, $\mu=0.5$, $p=0.5$, $q=0.5$, and $l=10$. A random rewiring strategy is applied, but similar dynamics are observed with different strategies.}
\label{fig:coevo}
\end{figure}

To illustrate the dynamics of the model, in Figure~\ref{fig:coevo} we show one simulation run. Over time, due to social influence and unfriending, each user is increasingly exposed to similar messages endorsed by friends (Fig.~\ref{fig:coevo}A), and the system reaches a steady state characterized by two distinctive features often observed in reality: opinion polarization (Fig.~\ref{fig:coevo}B) and network segregation (Fig.~\ref{fig:coevo}C). Note that by ``polarization'' we mean a division of opinions into distinct homogeneous groups, which are not necessarily at the extremes of the opinion range. 

\begin{figure}
\centering
\includegraphics[width=0.8\textwidth]{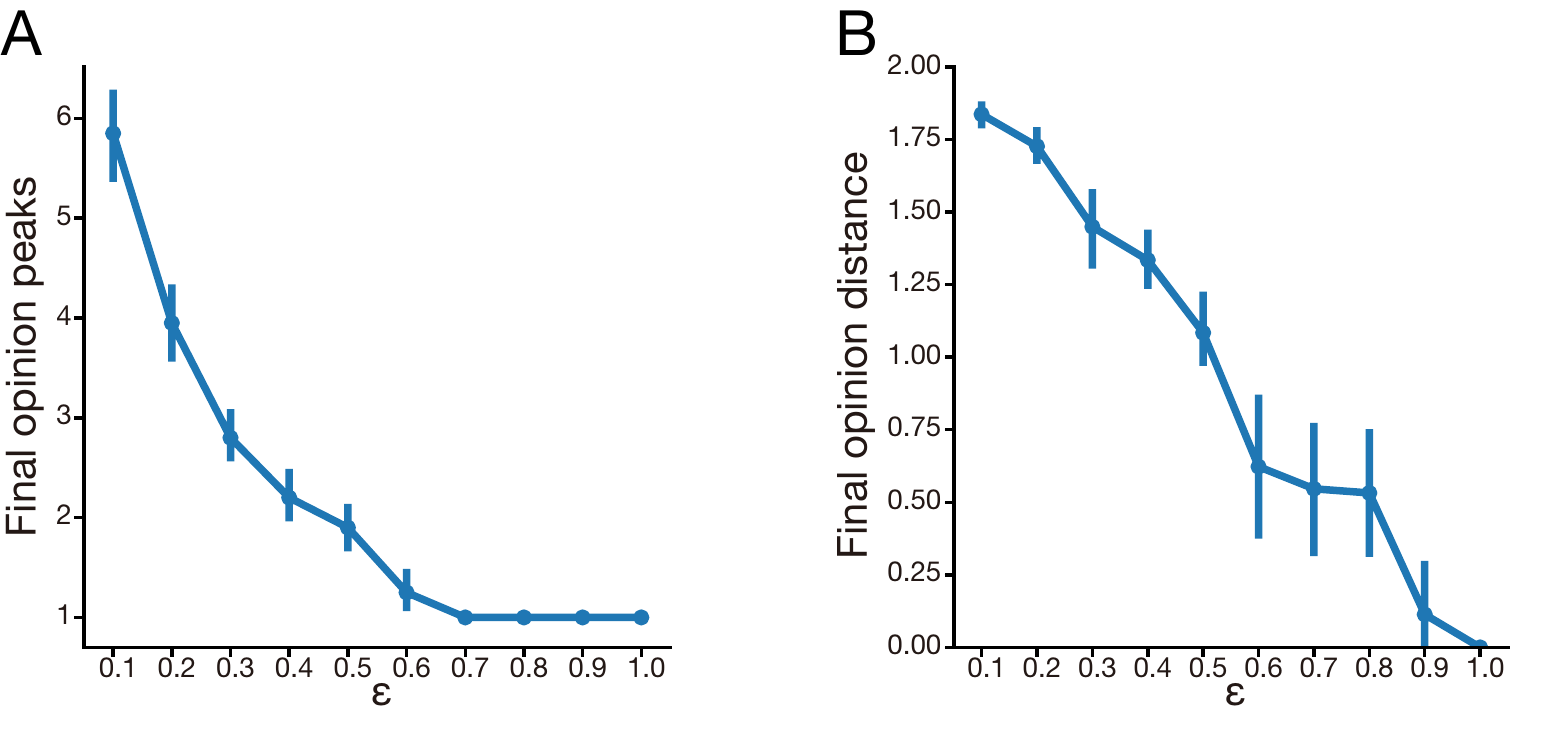}
\caption{Dependence of stationary opinions on bounded confidence parameter $\epsilon$: (A)~number of opinion peaks and (B)~maximum distance between opinions~\cite{peakutils}. The plots consider opinions at the steady state and show averages and standard deviations across 20 simulation runs with $N=100$, $E=400$, $\mu=0.5$, $p=0.5$, and $q=0.5$.}
\label{fig:epdep}
\end{figure}

We wish to examine the conditions under which opinion polarization and network segregation coevolve. Recall that our model has two mechanisms that appear to be relevant to this process: social influence (regulated by parameters $\epsilon$ and $\mu$) and rewiring (regulated by $q$). 
Let us first examine the role of the confidence bound parameter $\epsilon$. This threshold affects the number of final opinion clusters and the diversity of surviving opinions, in a manner consistent with the original bounded confidence model~\cite{Deffuant2000} and some of its extensions~\cite{DelVicario2017}. As shown in Fig.~\ref{fig:epdep}(A,B), the smaller $\epsilon$, the more opinion clusters with more heterogeneous opinions. If $\epsilon$ is large enough, the network converges to a single homogeneous opinion cluster. 

\subsection*{Role of Influence and Rewiring}

\begin{figure}
\centerline{\includegraphics[width=1.0\textwidth]{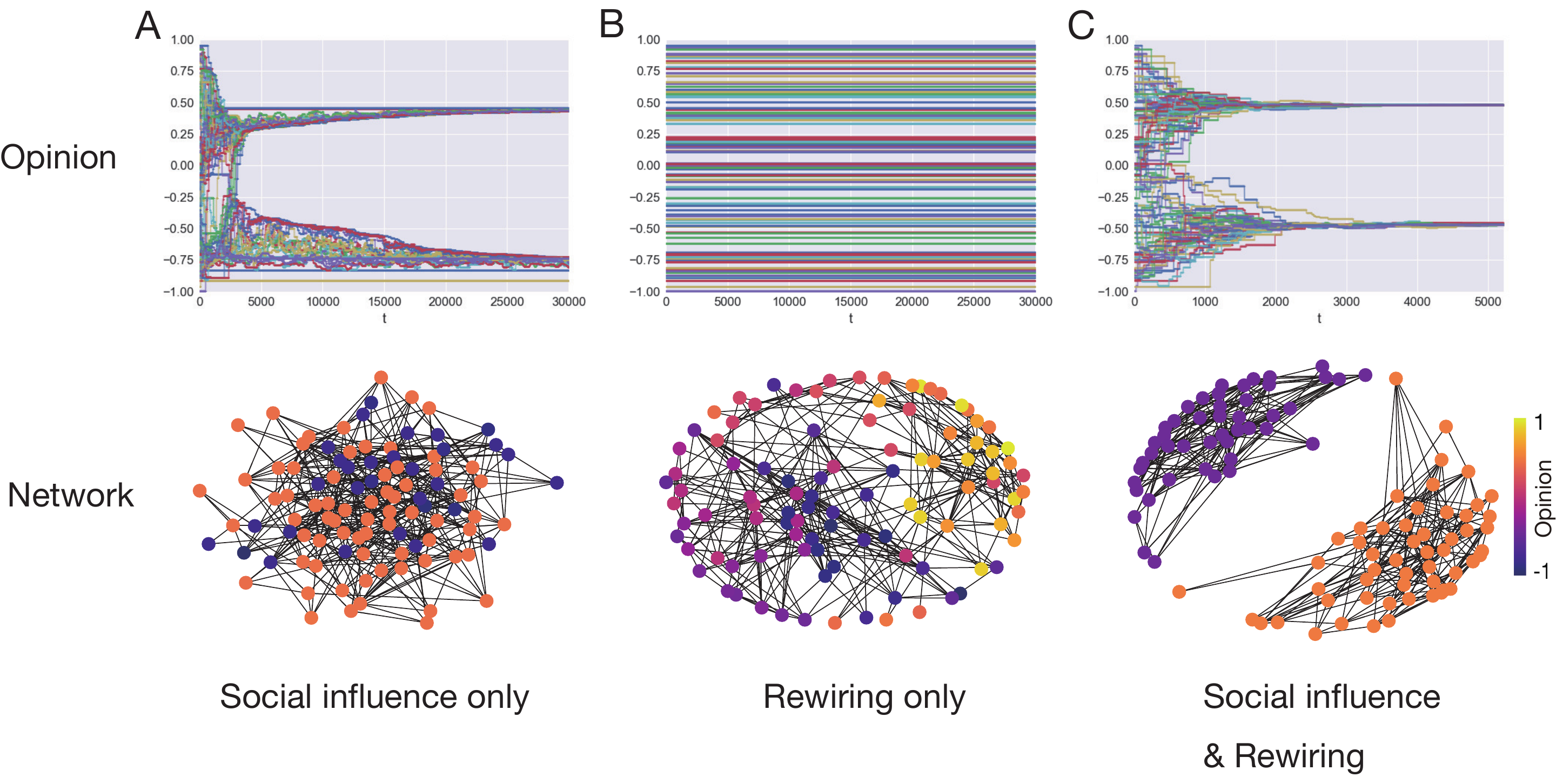}}
\caption{Conditions for the coevolution of opinion polarization (top) and network segregation (bottom). Left: $\mu=0.1$ and $q=0$. Center: $\mu=0$ and $q=0.1$. Right: $\mu=0.1$ and $q=0.1$.}
\label{fig:conditions}
\end{figure}

Next, let us explore the joint effects of  influence and rewiring. Here we limit our attention to the case $\epsilon = 0.4$, which yields, on average, two segregated opinion groups as illustrated in Fig.~\ref{fig:coevo}. In the presence of social influence alone without rewiring (Fig.~\ref{fig:conditions}A), the network structure is unaffected, but opinions may become polarized after a long time. In the presence of rewiring alone (Fig.~\ref{fig:conditions}B), no opinion change can happen but like-minded users cluster together making polarization inevitable; the network may become segregated after a very long time.  The joint effect of social influence and rewiring accelerates the joint emergence of both polarization and segregation (Fig.~\ref{fig:conditions}C). 

\begin{figure}
\centerline{\includegraphics[width=1.0\textwidth]{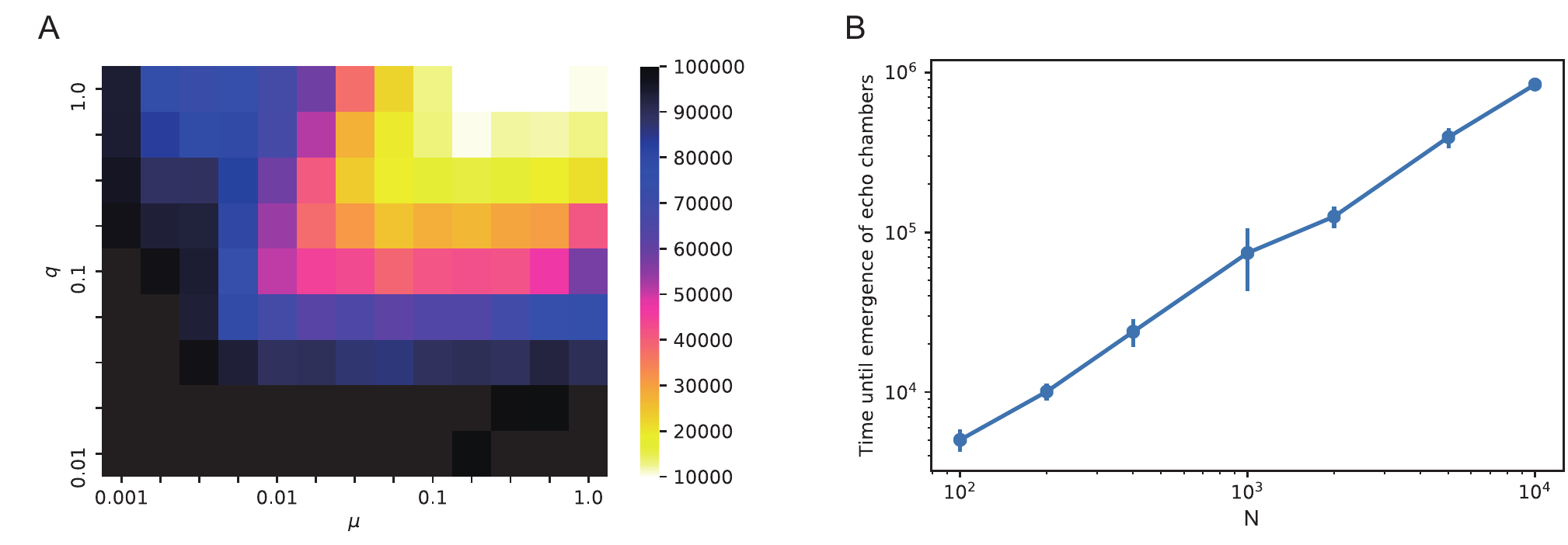}}
\caption{
Time until emergence of echo chambers as a function of (A) influence strength $\mu$ and rewiring rate $q$ and (B) network size ($N$). In (A), we use a logarithmic scale to explore small parameter values. For each $(\mu, q)$ parameter configuration we ran 200 simulations with $N=100$, $E=400$, $\epsilon = 0.4$, and $p=0.5$. A few simulations were excluded (see text), so that the median number of runs is 197. Colors represent averages across these simulation runs. The simulations were stopped after $t_{\text{max}}=10^5$ steps in cases when segregation and convergence have not both occurred yet. In (B), we use the same network density as in (A) and plot means and standard deviations across 10 simulation runs.}
\label{fig:mu_q_plot}
\end{figure}

To further explore how influence and rewiring jointly affect the speed of emergence of echo chambers, we plot in Fig.~\ref{fig:mu_q_plot}(A) the time until two conditions are both satisfied: (i)~the network clusters are segregated and (ii)~opinions are homogeneous, i.e., any two nodes within the same cluster have opinions that differ by less than the bounded confidence $\epsilon$. In some cases, a cluster may form that is smaller than the out-degree of one or more of its nodes, so that links from these nodes cannot be rewired to their own cluster; these cases are excluded because segregation can never occur. With these exceptions, segregated clusters always form when $q>0$ and $\mu>0$. This may take a long time if $q$ and $\mu$ are exceedingly small. However, even relatively small amounts of influence and rewiring greatly accelerate the emergence of segregated and polarized echo chambers. When both the rewiring rate $q$ and the influence strength $\mu$ are above 0.1, echo chambers appear in a fraction of the time. We therefore observe a synergistic effect in which influence and unfollowing reinforce each other in leading to the formation of echo chambers. 
Time until the emergence of echo chambers also shows an expected linear dependence on network size, as shown in Fig.~\ref{fig:mu_q_plot}(B).

\subsection*{Effects of Rewiring Strategies}

All three rewiring strategies used in the model (random, repost, recommendation) produce comparable stable states in terms of the number and diversity of stationary opinion clusters. In other words, what leads to an echo chamber state is selective unfollowing and not the specific mechanism by which one selects a new friend to follow. However, the emergence of echo chambers is greatly accelerated by the rewiring strategy based on recommendations of users with concordant opinions. The speed of convergence to the steady state is more than doubled compared to the other rewiring strategies.

\begin{figure}
\centerline{\includegraphics[width=1.0\textwidth]{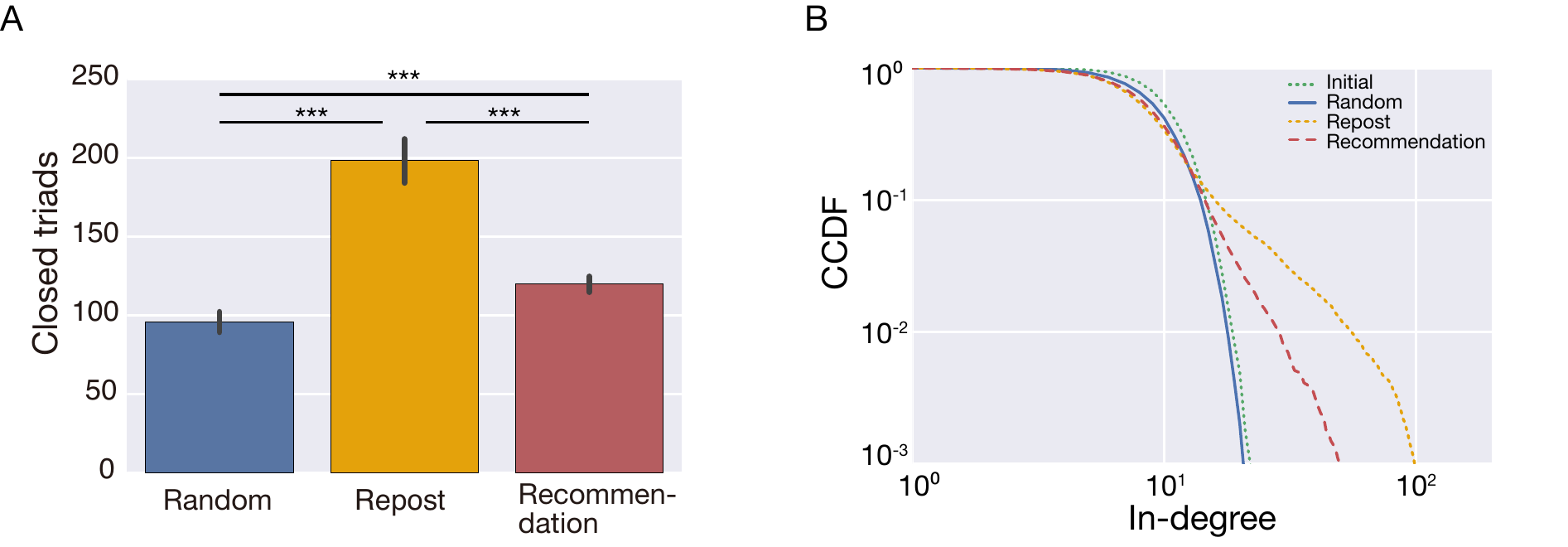}}
\caption{Effects of different rewiring strategies on evolved social networks. (A) The number of closed triads is averaged across 20 simulations with $N=100$ and $E=400$; all differences are statistically significant ($P<0.01$). (B) Cumulative in-degree distributions with $N=10^4$ and $E=10^5$.}
\label{fig:triads_degree}
\end{figure}

The rewiring strategy also affects the development of closed social triads. A closed social triad is a network motif with three nodes $A, B, C$ and links $A \rightarrow B, A \rightarrow C, C \rightarrow B$. It can be thought as the smallest unit of an echo chamber network~\cite{Jasny2015}, since it enables the same information to be transmitted from a source $A$ to a recipient $B$ through different paths, directly and via an intermediary $C$. As shown in Fig.~\ref{fig:triads_degree}A, rewiring strategies based on recommendations of users with concordant opinions and on exposure via reposts --- both common mechanisms in social media --- result in significantly more closed triads than following users at random. Repost-based rewiring, in particular, leads to doubling the number of directed closed triads, making it much more likely that users are exposed to the same opinions from multiple sources. The number of users posting/reposting a message can affect its ranking and be displayed to the user through platform-dependent interface elements, boosting user attention and exposure. 

Finally, the rewiring strategy affects the in-degree distribution of the network in the stable state. Compared to random rewiring, the other two methods yield more skewed in-degree distributions, indicating the spontaneous emergence of popular users with many followers, whose messages have potentially broader reach (Fig.~\ref{fig:triads_degree}B). Again, the effect on hubs is stronger in the case of repost-based rewiring. This is consistent with the \emph{copy model} for network growth, which approximates preferential attachment~\cite{kleinberg99b}. 
However, unlike the copy model, 
the number of nodes and links is fixed in our model; only the patterns of connection change. Thus, the skewed in-degree distribution arises due to the spread of information. 
Since recipients can see who originally posted each message, 
the originators of popular messages have the best chance of receiving new followers and becoming hubs. 

\subsection*{Empirical Validation}

It is tempting to use our model to reproduce a few stylized facts about empirical echo chambers. To this end, let us consider data about an empirical retweet network of US political conversations (see Methods A). To fit the model against this data, we plug in values of known parameters estimated in prior work, and then perform a sweep of the remaining parameters (Methods B). 
We simulate the resulting calibrated model to see if the synthetic network snapshot generated at the end of the simulation is in agreement with the observed snapshot of the empirical network (Methods C). As a stopping criterion for the simulations, we check that the simulated network has reached the same level of segregation as the empirical one (Methods D). 

\begin{figure}
    \centering
    \includegraphics[width=\textwidth]{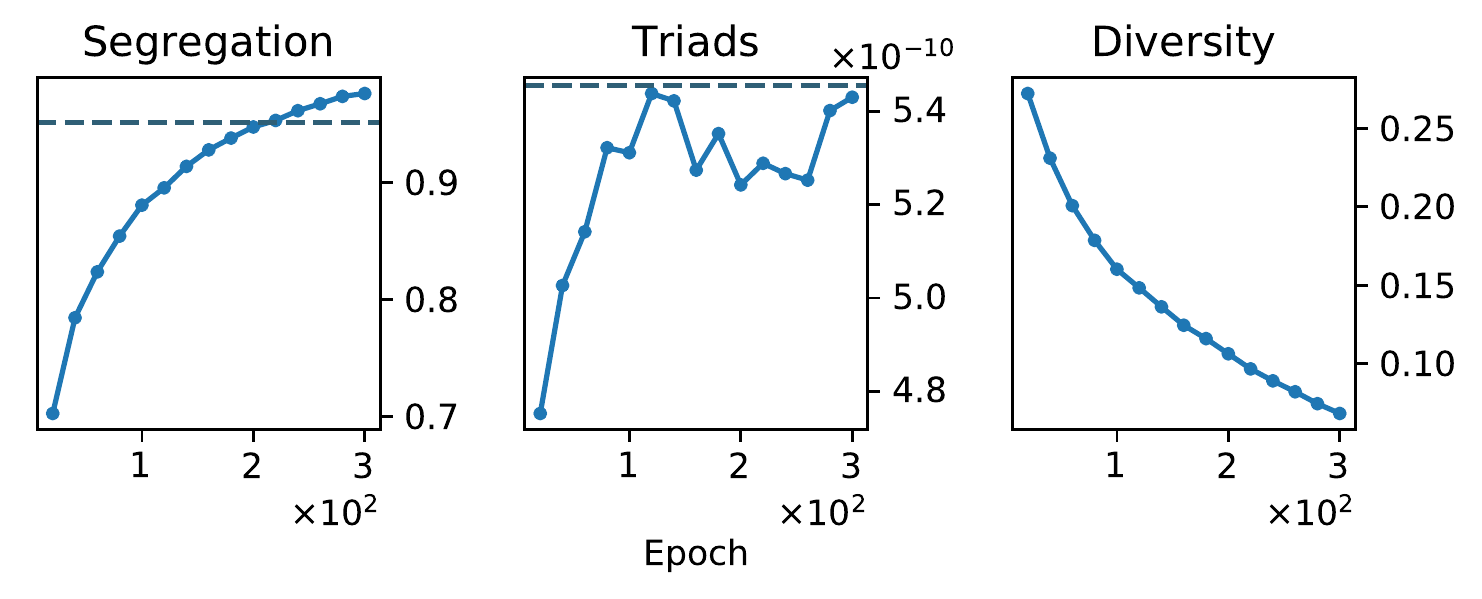}
    \caption{Comparison between model and empirical retweet networks. The solid blue lines represent the evolution of three metrics as a function of simulated time (epochs). The dashed line represents the empirical value for the segregation index (left) and fraction of closed directed triads (center), defined in Methods D and E respectively. Diversity (right) is the average distance between neighbor opinions.}
    \label{fig:validation}
\end{figure}

Fig.~\ref{fig:validation} shows the results. 
The segregation is one feature of the network structure that the model can reproduce. Aside from this, the random rewiring scheme used in the model cannot produce networks with heterogeneous degree. However we can draw a comparison between the empirical data and our simulations based on two other metrics. 
The first is the fraction of \emph{closed triads} in the network. To compute the number of triads, we record each time a user reposts something in our model as a `retweet,' and build a simulated retweet network. We then count all instances of closed directed triangles in both networks (Methods E). The central panel in Fig~\ref{fig:validation} shows that the fraction of triads in the synthetic network is consistent with that observed in the empirical data.

\begin{figure}[t]
    \centering
    \includegraphics[width=0.67\textwidth]{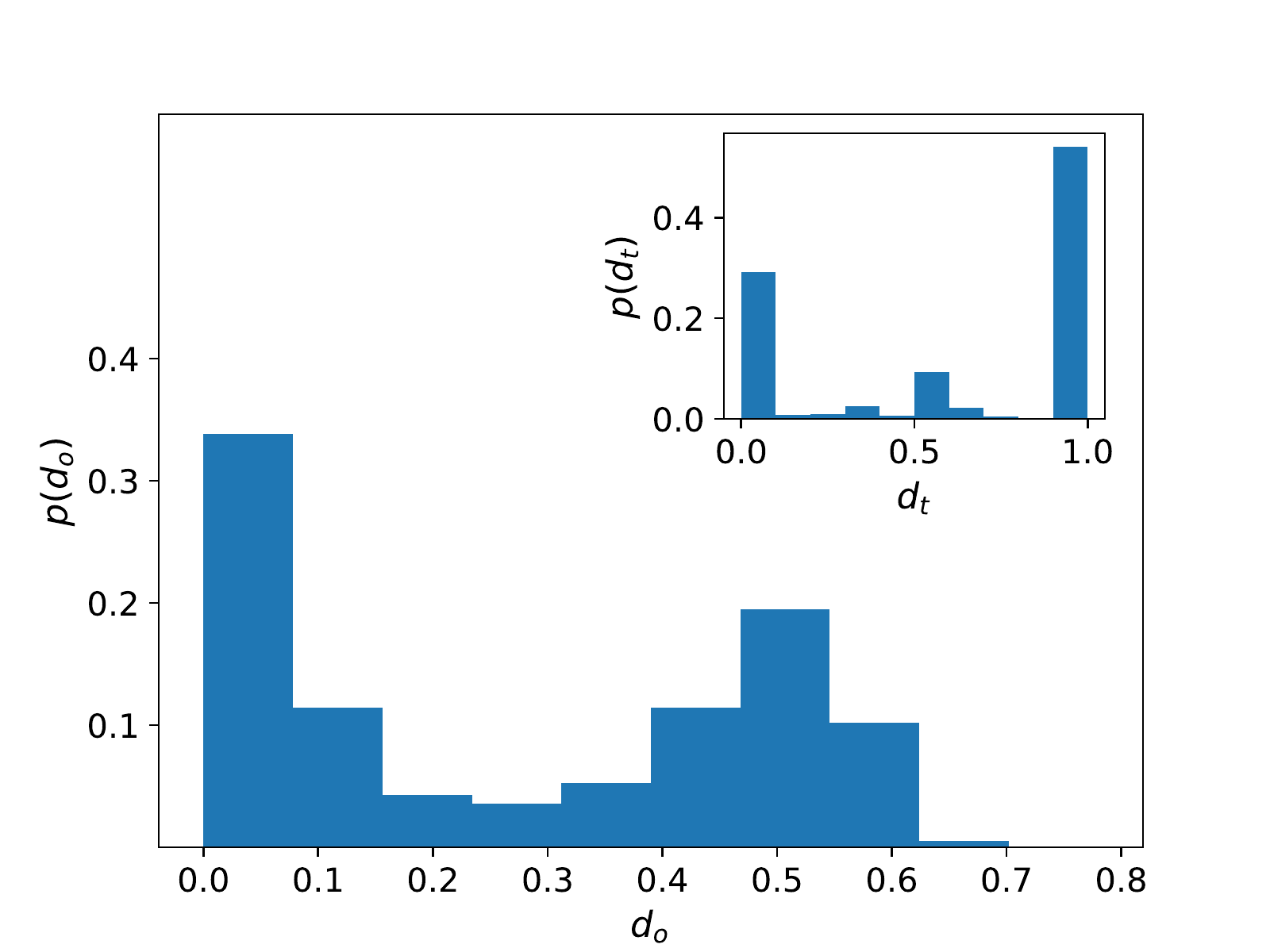}
    \caption{Distribution of pairwise opinion distances. The calibrated model was simulated until the segregation of the synthetic retweet network matched the observed one (see Methods E and Fig.~\ref{fig:validation}). The main plot shows the distribution of opinion distances $d_o$ across pairs of users in the simulated network. The inset shows the distribution of opinion distances $d_t$ among Twitter users from the empirical data.}
    \label{fig:opinions}
\end{figure}

The second metric is the distribution of \emph{opinion distances}. We infer the latent opinions of the Twitter users in our data from their hashtag usage (Methods F), and define a distance $d_t$ in hashtag binary vector space. In the model, we simply consider the distance $d_o(o_i, o_j) = | o_i - o_j |$ between two users in the opinion space.
Fig.~\ref{fig:validation} shows that the average distance between neighbor opinions decreases, but we also want to compare the distributions. 
Fig.~\ref{fig:opinions} shows that both distance distributions have peaks around low values of distance for users in the same cluster and around high values for users in different clusters. While the ways in which the distances are measured and consequently the distributions are necessarily different, the qualitatively similar bimodal behaviors suggest that the calibrated model attains an analogous level of opinion polarization in correspondence of the level of network segregation observed in the empirical data. 

\begin{figure}
    \centering
    \includegraphics[width=\textwidth]{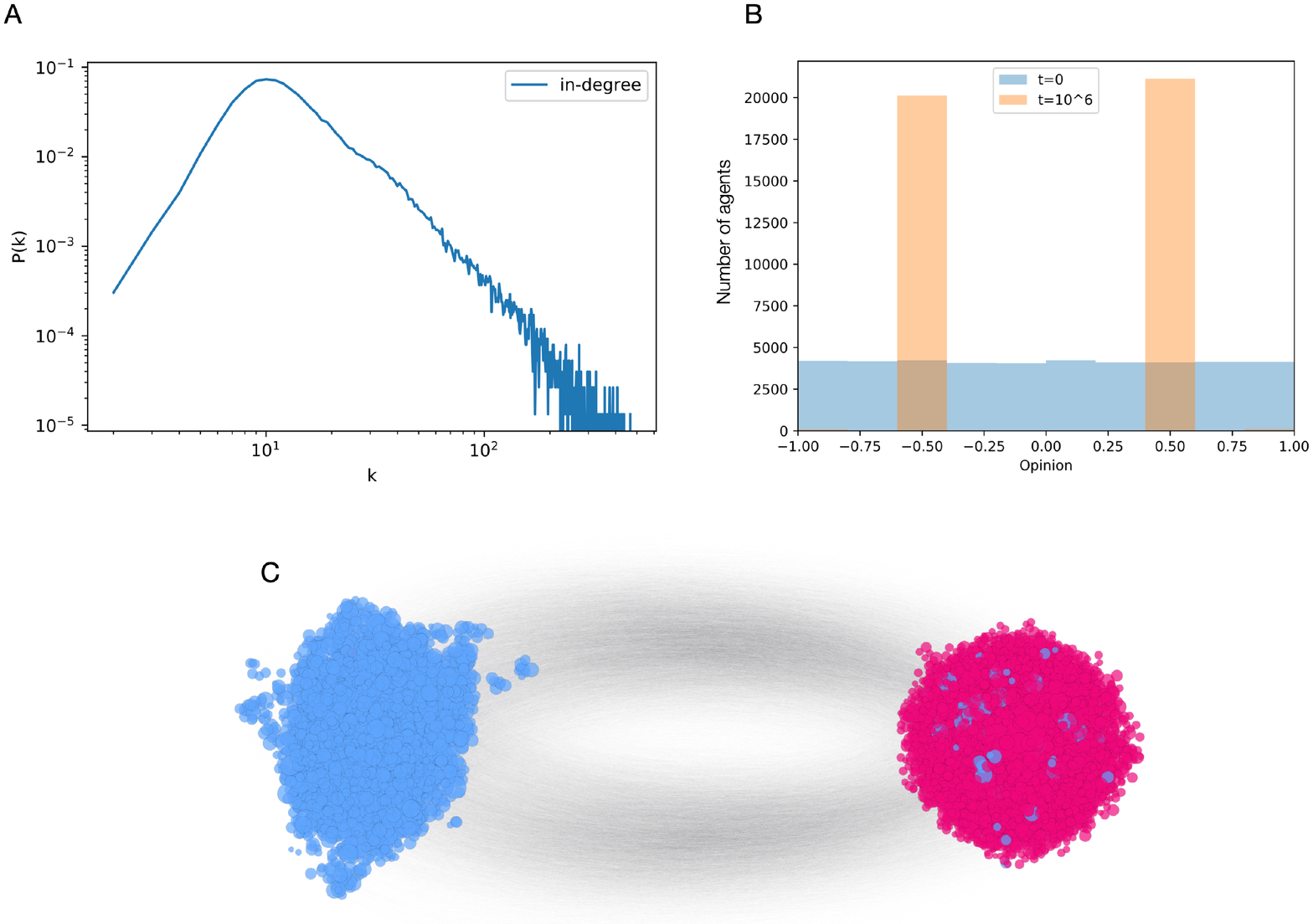}
    \caption{Simulation of model on a large empirical network. 
    (A)~Heavy-tailed in-degree distribution in the initial network. 
    (B)~Initial ($t=0$) and final ($t=10^6$) opinion distributions, showing the emergence of polarization. (C)~Visualization of the follower network after $t=10^6$ steps, showing strong segregation.}
    \label{fig:snap}
\end{figure}

To confirm that the proposed model yields echo chambers when applied to larger networks with realistic degree distributions, we conducted a simulation using an empirical Twitter follower network with $N=14{,}818$ nodes and $E=428{,}557$ edges (Methods A).
As shown in Fig.~\ref{fig:snap}, the simulation results confirm that both opinion polarization and network segregation emerge from this empirical network.

\section*{Discussion}

In studying an idealized social media platform using an agent-based model, we followed the rich tradition of several models of opinion dynamics under social influence~\cite{Castellano2009}, in which agents adjust their opinions based on those of the ones with whom they are connected (social influence), and rewire their ties with peers based on their shared opinions (social selection). Other models have explored the essential tension between social influence and social selection~\cite{Holme2006,Crandall2008,durrett2012graph,yu2017opinion,Teza:2018:Network}. The effect of the interaction between these two mechanisms on the emergence of opinion clusters has also been studied within a bounded confidence framework similar to the one presented here~\cite{PhysRevE.77.016102,Kozma_2008}. 

All of these models present extremely simplified versions of real social networks. While the model presented here is no exception, it seeks to capture more closely the key components of social media by focusing on indirect interactions enabled by information diffusion, in addition to disagreement-driven dissolution of ties via unfollowing/unfriending. Furthermore, our model combines social influence and selection with the competition for limited attention, which has been shown to explain the empirical scale-free distribution of content popularity in social media~\cite{Weng2012,Gleeson2014}. 

The results presented here suggest that the proliferation of online echo chambers may be an inevitable outcome of basic cognitive and social processes facilitated by social media: namely, the human tendency to be influenced by information and opinions to which one is exposed, and that of disliking disagreeable social ties. Social influence and rewiring appear to provide synergistic conditions for the rapid formation of completely segregated and polarized echo chambers; this phenomenon is accelerated by an order of magnitude in the presence of both relatively strong influence and relatively common unfollowing, compared to cases when either mechanism is weak. 

A social network that is both segregated and polarized can be also generated with a variant of the Schelling model~\cite{schelling1971dynamic} on networks, proposed by Henry~\textit{et al.}~\cite{Henry2011}. This model is based on aversion-driven rewiring, but starts from a bimodal distribution of opinions. Our approach shows how both polarization and segregation emerge without assuming that opinions are already polarized. Another variation of the the Schelling model has been recently used to show that homophilous rewiring inevitably leads to fragmented echo chambers~\cite{blex2020positive}.

The literature has explored other factors and mechanisms that foster the emergence of isolated cultural or political subnetworks as well as polarization of opinions. Network transitivity, which is also incorporated in our model, has been shown to lead to the formation of groups when combined with reciprocity~\cite{doi:10.1177/0956797614521816} or with one-to-many communication~\cite{Keijzer2018}. Pressures toward stronger opinions~\cite{doi:10.1098/rsos.181122} or more radical opinions~\cite{PhysRevLett.124.048301} are not included in our model, nor are repulsion effects driving opinions farther apart~\cite{DelVicario2016a,DelVicario2017}. Echo chambers can also emerge from cognitive mechanisms, such as confirmation bias, when information propagates through centralized channels reaching a large portion of the population~\cite{Geschke2018}. Finally, geographic segregation in urban areas may promote polarization in both physical and online spaces by fracturing the social space of mutual exposure~\cite{doi:10.1098/rsos.190573}. Our model focuses exclusively on information spreading mechanisms that are characteristic of online social media platforms.

Focusing on the rewiring of social ties, we tested three different selection mechanisms: two inspired by triadic closure and social recommendation --- intended to model the ways in which social media work in practice --- and one based on purely random choice. All variations yield qualitatively similar steady states, suggesting that disagreement-driven unfollowing is a sufficient rewiring condition for echo chamber emergence. The number of groups does depend on the tolerance to different opinions, as predicted by bounded confidence models~\cite{Deffuant2000}. This suggests that the extent to which social media allow users to exercise their preferences in determining their connections will either integrate or fragment interaction~\cite{van2005global}.

The more realistic selection mechanisms help explain two additional features of online social networks. First, the presence of users with many followers. These hub nodes affect the dissemination of the same messages in many cases, but not always~\cite{bakshy2011everyones}. Second, the large number of closed triads~\cite{weng2013role}. Triadic closure connects individuals to friends of their friends, facilitating repeated exposure to the same opinion. Such ``echoes'' are a powerful reinforcing mechanism for the adoption of beliefs and behaviors~\cite{Centola2007}.  

Although our model does not account for the adoption of false information, it has been speculated that echo chambers may make social media users more vulnerable to this kind of manipulation~\cite{lazer2018science,LEWANDOWSKY2017353}. There are multiple ways in which echo-chamber structure may contribute to the spread of misinformation. 
First, because people are repeatedly exposed to homogeneous information inside an echo chamber, the selection of belief-consistent information and the avoidance of belief-inconsistent information are facilitated, reinforcing confidence in minority opinions, such as conspiracy theories and fabricated news, even in the presence of preponderant contrary evidence~\cite{HillsProliferation18}. 
Second, echo chambers foster herding, which may lead to quick and premature convergence to suboptimal solutions of complex problems and simplistic interpretations of complex issues~\cite{Nematzadeh2017popularity-bias,HillsProliferation18}. 
Third, the threshold for perceiving a piece of content as novel may be lower within echo chambers by virtue of the reduced diversity of viewpoints to which people are exposed. The crafting of false news with perceived novelty may thus be promoted, leading to faster and broader consumption of misinformation~\cite{vosoughi2018spread} and triggering the production of more information about similar topics~\cite{ciampaglia2015production}. 
Finally, echo chambers may reinforce the influence bias of personalized filtering algorithms toward a user's current opinions~\cite{Perra2019personalisation}. 
On the other hand, recent experimental results suggest that social information exchange in homogeneous networks increases accuracy and reduces polarization~\cite{Becker201817195}, casting doubt on theories that political echo chambers reduce belief accuracy. More work is certainly needed to understand the relationship between online echo chambers and misinformation.

Our results suggest possible mitigation strategies against the emergence of echo chambers on social media. Often-recommended solutions involve exposure to content that increases a user's social distance from their preferences. However, such strategies must be consistent with current understanding of cognitive biases~\cite{HillsProliferation18}. For example, it does not help to promote content that will be disregarded~\cite{LEWANDOWSKY2017353}. A more neutral possibility suggested by our findings (Fig.~\ref{fig:triads_degree}A) is to discourage triadic closure when recommending the formation of new social ties. By exposing the sources of information reposted by our friends and recommending similar users, social media platforms encourage linking choices that maximize triadic closure. While exact knowledge about recommendation algorithms is not available, we know that triadic closure plays a key role. But even among the many friends of our friends, an algorithm could optimize for diversity of opinions. Moreover, the complete dissolution of ties with those users with whom one disagrees should be discouraged, for example by providing alternative mechanisms --- some social media platforms are experimenting with solutions like snooze buttons~\cite{NewsfeedFYI} --- or the possibility to block only certain types of information, but not others. Another approach is to alert users who are about to unfollow their only conduits to certain types or sources of information. 

As we better understand the unintended consequences of social media mechanisms, ethical and transparent user studies are needed to carefully evaluate countermeasures before they are deployed by platforms. We must not only ensure that new mechanisms mitigate undesired outcomes, but also that they do not create new vulnerabilities.


\section*{Methods}
\label{methods}

\subsection*{A. Data}
\label{methods:data}

To evaluate the model's prediction, we use empirical data from Conover \textit{et al.}~\cite{Conover2011}, who studied the political polarization of Twitter users. The data comprises a sample of public tweets about US politics, collected during the 6 weeks prior to the 2010 US midterm elections. The tweets were obtained from 
a 10\% random sample of all public tweets. 
The dataset is available online at \url{cnets.indiana.edu/groups/nan/nan-datasets-and-data-tools/#icswm2011_2}. 

Tweets with hashtags about US politics were included in the dataset. The hashtags were drawn from a list, which was obtained by expanding a manually curated seed set of then-popular political hashtags, such as \texttt{\#TCOT} (`Top conservatives on Twitter') and \texttt{\#P2} (`Progressives 2.0'). This initial set was recursively expanded with co-occurring hashtags above a minimum frequency, until no additional hashtag could be found. Finally, the list was manually checked and any hashtag that was not about US politics was expunged. The final list included 6,372 hashtags about US politics.

Three networks are provided in the dataset: retweets, mentions, and retweets plus mentions combined. We used the largest strongly connected component of the retweet network ($N = 18,470$ and $E = 48,365$), which is known to be polarized in two groups, roughly corresponding to the two main US political factions --- conservatives and progressives. The network is the same shown in Fig.~\ref{fig:echocham}.

For the larger empirical network, we started from a follower network with $N = 41{,}652{,}230$ nodes and $E = 1{,}468{,}364{,}884$ edges~\cite{snapnets}. To produce a more manageable dataset, we first randomly sampled edges from the graph and focused on the network spanned by these edges, and then applied  $k$-core decomposition~\cite{alvarez2006large} with $k = 30$. This yielded a reduced network with $N=14{,}818$ nodes and $E=428{,}557$ edges, which was used to initialize the simulation.

\subsection*{B. Parameter Fitting}
\label{methods:fitting}

Our model includes several parameters that need to be estimated. The rate of reposting was set to $p = 0.25$ based on empirical results from Twitter~\cite{qiu2017limited}. Similarly, the screen length was set to $l = 10$, which is close to the average number of times that a user stops on a post during a scrolling session on a social blogging platform~\cite{qiu2017limited}.

The number of nodes in the simulations was taken to be the same as the number of Twitter users in the empirical retweet network. Edges were drawn at random between any two nodes with probability chosen so that the density of the follower graph is $d=1.8 \times 10^{-4}$. This value is within the range observed in the literature~\cite{cha2010measuring,bollen2011happiness}. 

We performed a parameter scan for the rest of the parameters, finding the following values: influence strength $\mu = 0.015$ and tolerance $\epsilon = 0.65$. Note that the tolerance value to reproduce the two opinion clusters in the US-based online political conversations is larger than the value $\epsilon = 0.4$ found for smaller networks (Fig.~\ref{fig:epdep}). Finally, for simplicity, we use the random rewiring rule.

\subsection*{C. Model Evaluation}
\label{methods:validation}

Our goal is to compare model predictions about the emergence of echo chambers with empirical data from social media. Unfortunately, lacking a probability distribution over the data, our model does not allow us to compute the likelihood of a given network or distribution of opinions. Thus we need to devise a method to evaluate our approach. This has become a common challenge, especially with the rise of agent-based modeling in the social sciences~\cite{epstein1996growing}. There is a vast literature devoted to developing rigorous methods to test simulation models on empirical data of social phenomena~\cite{windrum2007,Ciampaglia2013}. Although no single universal recipe exists, we adopt the common approach of generating synthetic data from our agent-based model and comparing them to the empirical data under appropriate distance measures.

Our main hypothesis is that both social influence and rewiring are required to reproduce patterns consistent with the empirical data. Under those conditions, the system will always reach a state in which there will be no ties connecting users with discordant opinions (see Fig.~\ref{fig:conditions}). However, the empirical network is not completely disconnected in two communities. Therefore, we simulate our model until the system reaches the same level of segregation observed in the empirical data, and compare the two networks. 

The empirical and model networks are however different. The former is a network of retweets, while the latter is a network of `follower' ties. Therefore we cannot compare these two networks directly, but instead we generate a synthetic retweet network from the simulated data. Every time a user performs a `repost' action in our simulations, we count it as a retweet, and add the corresponding edge in the simulated retweet network. 

The plots in Figs.~\ref{fig:validation} and \ref{fig:opinions} are based on snapshots of this synthetic retweet network, taken at evenly spaced time  intervals of 10 epochs each. Each epoch consists of $N$ steps of the model, so that each user performs one post and/or rewiring action per epoch on average. At the end of each tenth epoch we consider the latest $E$ distinct retweet edges, so that each simulated network snapshot is guaranteed to have the same number of edges as the empirical one. We then consider the largest strongly connected component of each simulated network snapshot. Therefore, the two networks do not generally match in the number of nodes.

\subsection*{D. Segregation}
\label{methods:segregation}

To measure the segregation in both the simulated and empirical networks, we group users into two clusters $(C_-, C_+)$. In the simulated network, $C_-$ is defined as the set of users having opinion $o < 0$ and $C_+$ as the set with $o \ge 0$. In the empirical network, the two clusters correspond to the labels obtained via label propagation~\cite{Conover2010predicting}. Let $E_b$ denote the set of edges connecting nodes in different clusters. We define the \emph{segregation index} as: 
\begin{equation}
    s = 1 - \frac{| E_b |}{2d \, | C_+ | \, | C_- |}.
\end{equation}
The segregation index compares the actual number of edges across the two clusters with the number we would observe in a random network with the same density $d$. When the network is completely segregated, $s=1$.

\subsection*{E. Closed Triads}
\label{methods:triads}

Let us denote by $i \rightarrow j$ a directed edge from node $i$ to node $j$.
A \emph{closed directed triangle} or \emph{closed triad} is any node triplet $i, j, k \in V$ such that $\{i \rightarrow j, j \rightarrow k, i \rightarrow k\} \subseteq E$. An \emph{open directed triangle} or \emph{open triad} is any node triplet for whom only a proper subset of those edges exists in $E$. Let us denote by $T$ the set of closed triads and by $T_u$ the set of open triads. We compute the frequency of closed triads as the ratio  
\[
 f_T = \frac{|T|}{|T \cup T_u|} = \frac{|T|}{N_T! / (N_T-3)!},
\]
where $N_T = \binom{N}{3}$ is the number of all node triplets.

\subsection*{F. Latent Opinion Inference}
\label{methods:opinions}

Our model generates opinions in the $[-1,+1]$ range, while the empirical network has binary labels (`liberal' or `conservative') inferred from a training set and propagation through the network~\cite{Conover2010predicting}. Comparing the opinions predicted by the model to these labels would not be meaningful, since the labels are trivially correlated with the network structure, by construction. 

A more meaningful comparison is between pairwise opinion distances, which we can generate for the Twitter users using a criterion that is not induced by the network's community structure. Since hashtag usage is also polarized~\cite{Conover2011}, we infer opinions distances from adopted hashtags. We say that a hashtag is adopted by a user if it is found either in tweets retweeted by the user (incoming edges), or in tweets by the user that were retweeted by someone else (outgoing edges). Let us consider the $i$-th user and the $k$-th hashtag. We define an \emph{empirical opinion vector} $\hat{o}_i\in \{0,1\}^{D}$ where $\hat{o}_{ik} = 1$ if user $i$ adopted hashtag $k$, and $0$ otherwise. 
We define the \emph{empirical opinion distance} between two opinion vectors based on shared tags:
\begin{equation}
d_t\left(\hat{o}_{i}, \hat{o}_{j}\right) = 1 - \frac{\hat{o}_{i} \cdot \hat{o}_{j}}{\min\left\{\lVert\hat{o}_{i}\rVert,\lVert\hat{o}_{j}\rVert\right\}},
\end{equation}
where $\lVert\cdot\rVert$ is the $\mathcal{L}_1$ norm, or number of shared tags. To mitigate the effects of sparsity and noise, we use only the $D=20$ most popular hashtags to define the vectors. The selected hashtags were adopted by 93\% of the users. We restrict the retweet network to the subgraph spanned by those users, but the overall results do not change significantly if we select enough hashtags to cover 100\% of the users.

\section*{Acknowledgements}

We are grateful to Thomas Hills and Marco Minici for helpful feedback. K.S. was supported by JST PRESTO grant no.~JPMJPR16D6, JST CREST grant no.~JPMJCR17A4, and JSPS/MEXT KAKENHI grant numbers JP19H04217 and JP17H06383 in \#4903. 
G.L.C. was supported by the Indiana University Network Science Institute, where he carried out the work presented here.
F.M. and A.F. were supported in part by DARPA contract no.~W911NF-17-C-0094.
Any opinions, findings, and conclusions or recommendations expressed in this material are those of the authors and do not necessarily reflect the views of the funding agencies.

\section*{Author Contributions}

K.S., G.L.C., A.F., and F.M. designed the research. K.S. and W.C. performed simulations, W.C. and F.M. analyzed the data. H.P. developed the online demo. K.S., G.L.C., F.M., and A.F. drafted the manuscript. All authors reviewed and approved the manuscript.

\newpage
\appendix

\section*{Appendix: Prevalence of Unfollow Events}

Our model features edge rewiring to represent the dissolution of a social tie. This is based on the assumption that users cut social links by occasionally unfollowing friends. To get a more quantitative estimate of empirical unfollow frequencies, we collected 451,844 tweets from April 25, 2020 to May 25, 2020 from a 10\% random sample of tweets. In this sample, 16,317 users had at least two tweets. We sorted the tweets chronologically, and for each pair of consecutive tweets by the same user we measured the change $\Delta f$ in the number of friends of the user (i.e., the number of people followed by the user). 
By comparing the timestamps of the two tweets we could estimate the change in the number of friends per day. Similarly, by comparing the total numbers of tweets, we estimated the change in the number of friends per tweet. In the latter case, we removed pairs where the number of tweets decreased due to tweet deletions. 
Finally, we averaged across all consecutive pairs from the same user to get a single average change estimate for each user.

\begin{figure}[b]
    \centering
    \includegraphics[width=\textwidth]{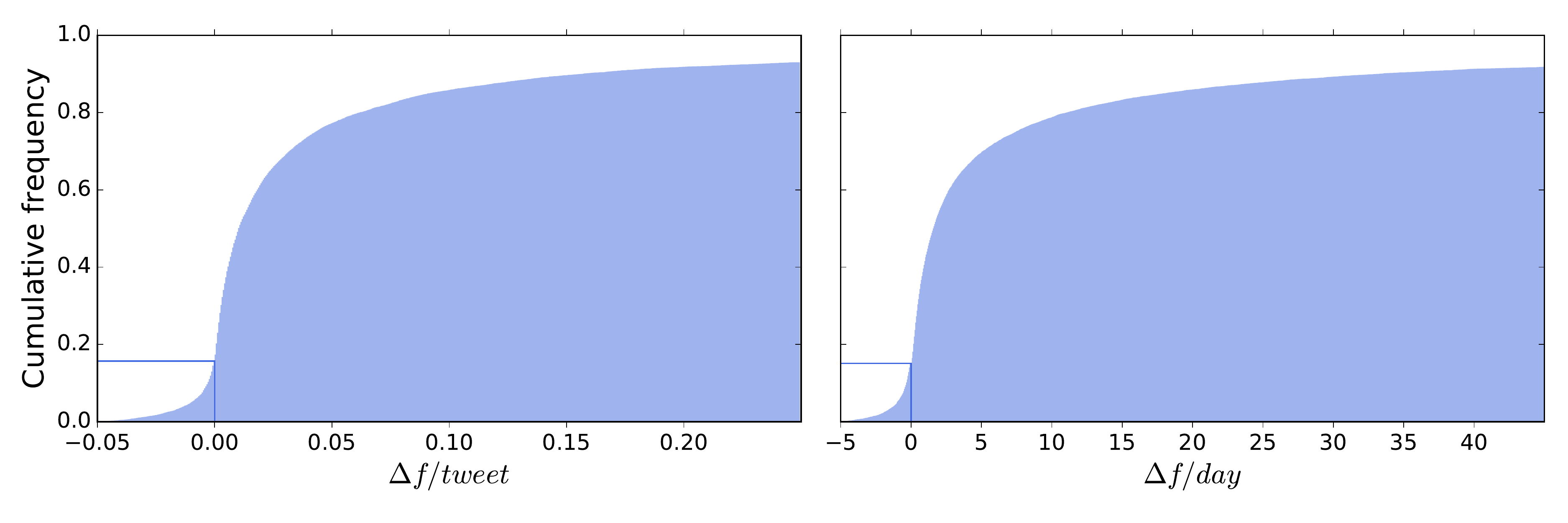}
    \caption{Distribution of average change in the number of user friends per tweet (left) and per day (right), based on a sample of active Twitter users.}
    \label{fig:unfollow}
\end{figure}

Fig.~\ref{fig:unfollow} plots the cumulative distributions of the average changes in numbers of followers across the active Twitter users in the sample. 
Approximately 18.5\% of users have negative changes, meaning that the number of friends has decreased --- more unfollows than new friends, on average.  This number can be understood as a lower bound on the actual probability of unfollowing.


\begin{thebibliography}{10}

\bibitem{Adamic2003}
Lada~A Adamic and Eytan Adar.
\newblock Friends and neighbors on the web.
\newblock {\em Social Networks}, 25(3):211--230, 2003.

\bibitem{alvarez2006large}
J~Ignacio Alvarez-Hamelin, Luca Dall'Asta, Alain Barrat, and Alessandro
  Vespignani.
\newblock Large scale networks fingerprinting and visualization using the
  k-core decomposition.
\newblock In {\em Advances in Neural Information Processing Systems}, pages
  41--50, 2006.

\bibitem{Backstrom2016}
Lars Backstrom.
\newblock Serving a billion personalized news feeds.
\newblock In {\em Proceedings of the Ninth ACM International Conference on Web
  Search and Data Mining}, WSDM '16, pages 469--469, New York, NY, USA, 2016.
  ACM.

\bibitem{Backstrom2006}
Lars Backstrom, Dan Huttenlocher, Jon Kleinberg, and Xiangyang Lan.
\newblock Group formation in large social networks: Membership, growth, and
  evolution.
\newblock In {\em Proceedings of the 12\textsuperscript{th} ACM SIGKDD
  International Conference on Knowledge Discovery and Data Mining}, KDD '06,
  pages 44--54, New York, NY, USA, 2006. ACM.

\bibitem{bakshy2011everyones}
Eytan Bakshy, Jake~M. Hofman, Winter~A. Mason, and Duncan~J. Watts.
\newblock Everyone's an influencer: Quantifying influence on twitter.
\newblock In {\em Proceedings of the Fourth ACM International Conference on Web
  Search and Data Mining}, WSDM '11, pages 65--74, New York, NY, USA, 2011.
  ACM.

\bibitem{Bakshy2015}
Eytan Bakshy, Solomon Messing, and Lada~A Adamic.
\newblock {Exposure to ideologically diverse news and opinion on Facebook.}
\newblock {\em Science}, 348(6239):1130--1132, 2015.

\bibitem{PhysRevLett.124.048301}
Fabian Baumann, Philipp Lorenz-Spreen, Igor~M. Sokolov, and Michele Starnini.
\newblock Modeling echo chambers and polarization dynamics in social networks.
\newblock {\em Phys. Rev. Lett.}, 124:048301, Jan 2020.

\bibitem{Becker201817195}
Joshua Becker, Ethan Porter, and Damon Centola.
\newblock The wisdom of partisan crowds.
\newblock {\em Proceedings of the National Academy of Sciences}, 2019.

\bibitem{benkler2006wealth}
Y.~Benkler.
\newblock {\em The Wealth of Networks: How Social Production Transforms Markets
  and Freedom}.
\newblock Yale University Press, New Haven, CT, USA, 2006.

\bibitem{blex2020positive}
Christian Blex and Taha Yasseri.
\newblock Positive algorithmic bias cannot stop fragmentation in homophilic
  social networks.
\newblock Preprint 2001.02878, arXiv, 2020.

\bibitem{bollen2011happiness}
Johan Bollen, Bruno Gon{\c{c}}alves, Guangchen Ruan, and Huina Mao.
\newblock Happiness is assortative in online social networks.
\newblock {\em Artificial Life}, 17(3):237--251, 2011.

\bibitem{Bonabeau2009}
Eric Bonabeau.
\newblock Decisions 2.0: the power of collective intelligence.
\newblock {\em MIT Sloan Management Review}, 50(2):45--52, Winter 2009.

\bibitem{Bright2016}
Jonathan Bright.
\newblock {Explaining the Emergence of Political Fragmentation on Social Media:
  The Role of Ideology and Extremism}.
\newblock {\em Journal of Computer-Mediated Communication}, 23(1):17--33, 2018.

\bibitem{Castellano2009}
Claudio Castellano, Santo Fortunato, and Vittorio Loreto.
\newblock Statistical physics of social dynamics.
\newblock {\em Rev. Mod. Phys.}, 81:591--646, May 2009.

\bibitem{centola_experimental_2011}
D.~Centola.
\newblock An {Experimental} {Study} of {Homophily} in the {Adoption} of
  {Health} {Behavior}.
\newblock {\em Science}, 334(6060):1269--1272, December 2011.

\bibitem{Centola2007}
Damon Centola and Michael Macy.
\newblock Complex contagions and the weakness of long ties.
\newblock {\em American Journal of Sociology}, 113(3):702--734, 2007.

\bibitem{cha2010measuring}
Meeyoung Cha, Hamed Haddadi, Fabricio Benevenuto, and Krishna Gummadi.
\newblock Measuring user influence in twitter: The million follower fallacy.
\newblock In {\em International AAAI Conference on Web and Social Media}, ICWSM
  '10, pages 10--17, Palo Alto, CA, USA, 2010. AAAI.

\bibitem{Ciampaglia2013}
Giovanni~Luca Ciampaglia.
\newblock A framework for the calibration of social simulation models.
\newblock {\em Advs Complex Sys.}, 16:1350030, 2013.

\bibitem{ciampaglia2015production}
Giovanni~Luca Ciampaglia, Alessandro Flammini, and Filippo Menczer.
\newblock The production of information in the attention economy.
\newblock {\em Scientific Reports}, 5:9452, May 2015.

\bibitem{Nematzadeh2017popularity-bias}
Giovanni~Luca Ciampaglia, Azadeh Nematzadeh, Filippo Menczer, and Alessandro
  Flammini.
\newblock How algorithmic popularity bias hinders or promotes quality.
\newblock {\em Scientific Reports}, 8:15951, 2018.

\bibitem{Conover2010predicting}
M~Conover, B~Gon\c{c}alves, J~Ratkiewicz, A~Flammini, and F~Menczer.
\newblock Predicting the political alignment of twitter users.
\newblock In {\em Proceedings of 3rd IEEE Conference on Social Computing
  (SocialCom)}, pages 192--199, 2011.

\bibitem{Conover2011}
Michael Conover, Jacob Ratkiewicz, Matthew Francisco, Bruno Gonçalves, Filippo
  Menczer, and Alessandro Flammini.
\newblock Political polarization on {T}witter.
\newblock In {\em International AAAI Conference on Web and Social Media}, ICWSM
  '11, pages 89--96, Palo Alto, CA, USA, 2011. AAAI.

\bibitem{Conover2012}
Michael~D Conover, Bruno Gon{\c c}alves, Alessandro Flammini, and Filippo
  Menczer.
\newblock {Partisan asymmetries in online political activity}.
\newblock {\em EPJ Data Sci.}, 1(1):6, 2012.

\bibitem{Crandall2008}
David Crandall, Dan Cosley, Daniel Huttenlocher, Jon Kleinberg, and Siddharth
  Suri.
\newblock Feedback effects between similarity and social influence in online
  communities.
\newblock In {\em Proceedings of the 14\textsuperscript{th} ACM SIGKDD
  International Conference on Knowledge Discovery and Data Mining}, KDD '08,
  pages 160--168, New York, NY, USA, 2008. ACM.

\bibitem{Deffuant2000}
g~deffuant, d~neau, and f~amblard.
\newblock {mixing beliefs among interacting agents}.
\newblock {\em adv. complex syst.}, 03(01n04):87--98, 2000.

\bibitem{DelVicario2016a}
Michela Del~Vicario, Alessandro Bessi, Fabiana Zollo, Fabio Petroni, Antonio
  Scala, Guido Caldarelli, H~Eugene Stanley, and Walter Quattrociocchi.
\newblock The spreading of misinformation online.
\newblock {\em Proc. Natl. Acad. Sci. U.S.A.}, 113(3):554--559, 2016.

\bibitem{DelVicario2017}
Michela Del~Vicario, Antonio Scala, Guido Caldarelli, H~Eugene Stanley, and
  Walter Quattrociocchi.
\newblock {Modeling confirmation bias and polarization}.
\newblock {\em Sci. Rep.}, 7:40391, 2017.

\bibitem{dubois2018echo}
Elizabeth Dubois and Grant Blank.
\newblock The echo chamber is overstated: the moderating effect of political
  interest and diverse media.
\newblock {\em Information, Communication \& Society}, 21(5):729--745, 2018.

\bibitem{durrett2012graph}
Richard Durrett, James~P. Gleeson, Alun~L. Lloyd, Peter~J. Mucha, Feng Shi,
  David Sivakoff, Joshua E.~S. Socolar, and Chris Varghese.
\newblock Graph fission in an evolving voter model.
\newblock {\em Proceedings of the National Academy of Sciences},
  109(10):3682--3687, 2012.

\bibitem{epstein1996growing}
J.M. Epstein, R.~Axtell, and 2050 Project.
\newblock {\em Growing Artificial Societies: Social Science from the Bottom
  Up}.
\newblock Complex Adaptive Systems. MIT Press, Cambridge, MA, USA, 1996.

\bibitem{doi:10.1098/rsos.181122}
Tucker Evans and Feng Fu.
\newblock Opinion formation on dynamic networks: identifying conditions for the
  emergence of partisan echo chambers.
\newblock {\em Royal Society Open Science}, 5(10):181122, 2018.

\bibitem{flache2017models}
Andreas Flache, Michael M{\"a}s, Thomas Feliciani, Edmund Chattoe-Brown,
  Guillaume Deffuant, Sylvie Huet, and Jan Lorenz.
\newblock Models of social influence: Towards the next frontiers.
\newblock {\em Journal of Artificial Societies and Social Simulation}, 20(4),
  2017.

\bibitem{flaxman2016filter}
Seth Flaxman, Sharad Goel, and Justin~M. Rao.
\newblock Filter bubbles, echo chambers, and online news consumption.
\newblock {\em Public Opinion Quarterly}, 80(S1):298--320, 2016.

\bibitem{fortunato2006topical}
S.~Fortunato, A.~Flammini, F.~Menczer, and A.~Vespignani.
\newblock Topical interests and the mitigation of search engine bias.
\newblock {\em Proceedings of the National Academy of Sciences},
  103(34):12684--12689, 2006.

\bibitem{friedkin2006structural}
Noah~E Friedkin.
\newblock {\em A structural theory of social influence}, volume~13.
\newblock Cambridge University Press, 2006.

\bibitem{garimella2018political}
Kiran Garimella, Gianmarco De~Francisci~Morales, Aristides Gionis, and Michael
  Mathioudakis.
\newblock Political discourse on social media: Echo chambers, gatekeepers, and
  the price of bipartisanship.
\newblock In {\em Proceedings of the 2018 World Wide Web Conference}, pages
  913--922, 2018.

\bibitem{Garrett2009echo}
R.~Kelly Garrett.
\newblock {Echo chambers online?: Politically motivated selective exposure
  among Internet news users}.
\newblock {\em Journal of Computer-Mediated Communication}, 14(2):265--285,
  2009.

\bibitem{Gentzkow2011}
Matthew Gentzkow and Jesse~M. Shapiro.
\newblock Ideological segregation online and offline.
\newblock {\em The Quarterly Journal of Economics}, 126(4):1799--1839, 2011.

\bibitem{Geschke2018}
Daniel Geschke, Jan Lorenz, and Peter Holtz.
\newblock The triple-filter bubble: Using agent-based modelling to test a
  meta-theoretical framework for the emergence of filter bubbles and echo
  chambers.
\newblock {\em British Journal of Social Psychology}, 58(1):129--149, 2019.

\bibitem{Gleeson2014}
James~P. Gleeson, Jonathan~A. Ward, Kevin~P. O'Sullivan, and William~T. Lee.
\newblock Competition-induced criticality in a model of meme popularity.
\newblock {\em Phys. Rev. Lett.}, 112:048701, January 2014.

\bibitem{doi:10.1177/0956797614521816}
Kurt Gray, David~G. Rand, Eyal Ert, Kevin Lewis, Steve Hershman, and Michael~I.
  Norton.
\newblock The emergence of ``us and them'' in 80 lines of code: Modeling group
  genesis in homogeneous populations.
\newblock {\em Psychological Science}, 25(4):982--990, 2014.

\bibitem{guess2018avoiding}
Andrew Guess, Benjamin Lyons, Brendan Nyhan, and Jason Reifler.
\newblock Avoiding the echo chamber about echo chambers: Why selective exposure
  to like-minded political news is less prevalent than you think.
\newblock Knight Foundation, 2018.

\bibitem{Guess-diet}
Andrew~M. Guess.
\newblock {(Almost) Everything in Moderation: New Evidence on Americans' Online
  Media Diets}.
\newblock Unpublished, 2018.

\bibitem{hart2009feeling}
William Hart, Dolores Albarrac{\'\i}n, Alice~H Eagly, Inge Brechan, Matthew~J
  Lindberg, and Lisa Merrill.
\newblock Feeling validated versus being correct: A meta-analysis of selective
  exposure to information.
\newblock {\em Psychological Bulletin}, 135(4):555--588, 2009.

\bibitem{Henry2011}
Adam~Douglas Henry, Pawe{\l} Pra{\l}at, and Cun-Quan Zhang.
\newblock Emergence of segregation in evolving social networks.
\newblock {\em Proc. Natl. Acad. Sci. U.S.A.}, 108(21):8605--8610, 2011.

\bibitem{HillsProliferation18}
Thomas~T. Hills.
\newblock The dark side of information proliferation.
\newblock {\em Perspectives on Psychological Science}, 14(3):323--330, 2019.

\bibitem{Holme2006}
Petter Holme and M.~E.~J. Newman.
\newblock Nonequilibrium phase transition in the coevolution of networks and
  opinions.
\newblock {\em Phys. Rev. E}, 74:056108, November 2006.

\bibitem{Jamieson2008}
K.H. Jamieson and J.N. Cappella.
\newblock {\em Echo Chamber: Rush Limbaugh and the Conservative Media
  Establishment}.
\newblock Oxford University Press, 2008.

\bibitem{Jasny2015}
Lorien Jasny, Joseph Waggle, and Dana~R Fisher.
\newblock {An empirical examination of echo chambers in US climate policy
  networks}.
\newblock {\em Nat. Clim. Change}, 5(8):782--786, 2015.

\bibitem{katz1998struggle}
James~E. Katz.
\newblock Struggle in cyberspace: Fact and friction on the world wide web.
\newblock {\em The ANNALS of the American Academy of Political and Social
  Science}, 560(1):194--199, 1998.

\bibitem{Keijzer2018}
Marijn~A. Keijzer, Michael M{\"a}s, and Andreas Flache.
\newblock Communication in online social networks fosters cultural isolation.
\newblock {\em Complexity}, 2018(9502872), 2018.

\bibitem{kleinberg99b}
Jon~M Kleinberg, Ravi Kumar, Prabhakar Raghavan, Sridhar Rajagopalan, and
  Andrew~S Tomkins.
\newblock The web as a graph: measurements, models, and methods.
\newblock In {\em Computing and Combinatorics}, pages 1--17. Springer, 1999.

\bibitem{PhysRevE.77.016102}
Balazs Kozma and Alain Barrat.
\newblock Consensus formation on adaptive networks.
\newblock {\em Phys. Rev. E}, 77:016102, Jan 2008.

\bibitem{Kozma_2008}
Balazs Kozma and Alain Barrat.
\newblock Consensus formation on coevolving networks: groups' formation and
  structure.
\newblock {\em Journal of Physics A: Mathematical and Theoretical},
  41(22):224020, May 2008.

\bibitem{10.1145/1978942.1979104}
Haewoon Kwak, Hyunwoo Chun, and Sue Moon.
\newblock Fragile online relationship: A first look at unfollow dynamics in
  twitter.
\newblock In {\em Proceedings of the SIGCHI Conference on Human Factors in
  Computing Systems}, CHI ’11, page 1091–1100, New York, NY, USA, 2011.
  Association for Computing Machinery.

\bibitem{lazer2018science}
David M.~J. Lazer, Matthew~A. Baum, Yochai Benkler, Adam~J. Berinsky, Kelly~M.
  Greenhill, Filippo Menczer, Miriam~J. Metzger, Brendan Nyhan, Gordon
  Pennycook, David Rothschild, Michael Schudson, Steven~A. Sloman, Cass~R.
  Sunstein, Emily~A. Thorson, Duncan~J. Watts, and Jonathan~L. Zittrain.
\newblock The science of fake news.
\newblock {\em Science}, 359(6380):1094--1096, 2018.

\bibitem{Leskovec2008}
Jure Leskovec, Lars Backstrom, Ravi Kumar, and Andrew Tomkins.
\newblock Microscopic evolution of social networks.
\newblock In {\em Proceedings of the 14\textsuperscript{th} ACM SIGKDD
  International Conference on Knowledge Discovery and Data Mining}, KDD '08,
  pages 462--470, New York, NY, USA, 2008. ACM.

\bibitem{snapnets}
Jure Leskovec and Andrej Krevl.
\newblock {SNAP Datasets}: {Stanford} large network dataset collection.
\newblock \url{http://snap.stanford.edu/data}, June 2014.

\bibitem{LEWANDOWSKY2017353}
Stephan Lewandowsky, Ullrich~K.H. Ecker, and John Cook.
\newblock {Beyond Misinformation: Understanding and Coping with the
  ``Post-Truth'' Era}.
\newblock {\em Journal of Applied Research in Memory and Cognition},
  6(4):353--369, 2017.

\bibitem{marsden_homogeneity_1988}
Peter~V. Marsden.
\newblock Homogeneity in {Confiding} {Relations}.
\newblock {\em Social Networks}, 10(1):57 -- 76, 1988.

\bibitem{McPherson2001}
Miller McPherson, Lynn Smith-Lovin, and James~M Cook.
\newblock Birds of a feather: Homophily in social networks.
\newblock {\em Annual Review of Sociology}, 27(1):415--444, 2001.

\bibitem{doi:10.1098/rsos.190573}
Alfredo~J. Morales, Xiaowen Dong, Yaneer Bar-Yam, and Alex
  ‘Sandy’~Pentland.
\newblock Segregation and polarization in urban areas.
\newblock {\em Royal Society Open Science}, 6(10):190573, 2019.

\bibitem{NewsfeedFYI}
Shruthi Muraleedharan.
\newblock Introducing snooze to give you more control of your news feed.
\newblock \url{https://newsroom.fb.com/news/2017/12/news-feed-fyi-snooze/},
  December 2017.
\newblock Last accessed 27 February 2019.

\bibitem{peakutils}
Lucas~Hermann Negri and Christophe Vestri.
\newblock lucashn/peakutils: v1.1.0.
\newblock Zenodo, 2017.

\bibitem{Nickerson1998}
Raymond~S. Nickerson.
\newblock Confirmation bias: A ubiquitous phenomenon in many guises.
\newblock {\em Review of General Psychology}, 2(2):175--220, 1998.

\bibitem{Nikolov2018biases}
Dimitar Nikolov, Mounia Lalmas, Alessandro Flammini, and Filippo Menczer.
\newblock Quantifying biases in online information exposure.
\newblock {\em Journal of the Association for Information Science and
  Technology}, 70(3):218--229, 2019.

\bibitem{nikolov2015measuring}
Dimitar Nikolov, Diego~F.M. Oliveira, Alessandro Flammini, and Filippo Menczer.
\newblock Measuring online social bubbles.
\newblock {\em PeerJ Computer Science}, 1:e38, December 2015.

\bibitem{page2008difference}
S.E. Page.
\newblock {\em The Difference: How the Power of Diversity Creates Better
  Groups, Firms, Schools, and Societies}.
\newblock Princeton University Press, Princeton, NJ, USA, 2008.

\bibitem{pariser2011filter}
Eli Pariser.
\newblock {\em The filter bubble: What the Internet is hiding from you}.
\newblock Penguin UK, 2011.

\bibitem{Perra2019personalisation}
Nicola Perra and Luis E.~C. Rocha.
\newblock Modelling opinion dynamics in the age of algorithmic personalisation.
\newblock {\em Scientific Reports}, 9:7261, 2019.

\bibitem{qiu2017limited}
Xiaoyan Qiu, Diego F.~M.~Oliveira, Alireza Sahami~Shirazi, Alessandro Flammini,
  and Filippo Menczer.
\newblock Limited individual attention and online virality of low-quality
  information.
\newblock {\em Nature Human Behaviour}, 1:0132--, June 2017.

\bibitem{raeder2011predictors}
Troy Raeder, Omar Lizardo, David Hachen, and Nitesh~V. Chawla.
\newblock Predictors of short-term decay of cell phone contacts in a large
  scale communication network.
\newblock {\em Social Networks}, 33(4):245--257, 2011.

\bibitem{salganik2006experimental}
Matthew~J Salganik, Peter~Sheridan Dodds, and Duncan~J Watts.
\newblock Experimental study of inequality and unpredictability in an
  artificial cultural market.
\newblock {\em Science}, 311(5762):854--856, 2006.

\bibitem{schelling1971dynamic}
Thomas~C. Schelling.
\newblock Dynamic models of segregation.
\newblock {\em The Journal of Mathematical Sociology}, 1(2):143--186, 1971.

\bibitem{Schmidt2017}
Ana~L Schmidt, Fabiana Zollo, Michela Del~Vicario, Alessandro Bessi, Antonio
  Scala, Guido Caldarelli, {HE} Stanley, and Walter Quattrociocchi.
\newblock Anatomy of news consumption on facebook.
\newblock {\em Proc. Natl. Acad. Sci. {U.S.A.}}, 114(12):3035--3039, 2017.

\bibitem{Sears1967}
David~O. Sears and Jonathan~L. Freedman.
\newblock Selective exposure to information: a critical review.
\newblock {\em Public Opinion Quarterly}, 31(2):194--213, 1967.

\bibitem{shore2016network}
Jesse Shore, Jiye Baek, and Chrysanthos Dellarocas.
\newblock Network structure and patterns of information diversity on twitter.
\newblock {\em MIS Quarterly}, 42(3):849--872, 2016.

\bibitem{Sibona2011}
C.~Sibona and S.~Walczak.
\newblock Unfriending on facebook: Friend request and online/offline behavior
  analysis.
\newblock In {\em 2011 44\textsuperscript{th} Hawaii International Conference
  on System Sciences}, pages 1--10, January 2011.

\bibitem{gerrymandering2019}
Alexander~J. Stewart, Mohsen Mosleh, Marina Diakonova, Antonio~A. Arechar,
  David~G. Rand, and Joshua~B. Plotkin.
\newblock Information gerrymandering and undemocratic decisions.
\newblock {\em Nature}, 573(7772):117--121, 2019.

\bibitem{stroud2010polarization}
Natalie~Jomini Stroud.
\newblock Polarization and partisan selective exposure.
\newblock {\em Journal of Communication}, 60(3):556--576, 2010.

\bibitem{sunstein2002republiccom}
Cass~R. Sunstein.
\newblock {\em Republic.com}.
\newblock Princeton University Press, Princeton, NJ, USA, reprint edition,
  2002.

\bibitem{sunstein2017republic}
Cass~R. Sunstein.
\newblock {\em \#Republic: Divided Democracy in the Age of Social Media}.
\newblock Princeton University Press, Princeton, NJ, USA, 2017.

\bibitem{surowiecki2005wisdom}
J.~Surowiecki.
\newblock {\em The Wisdom Of Crowds}.
\newblock Anchor Books. Anchor Books, New York, NY, USA, 2005.

\bibitem{Teza:2018:Network}
Gianluca Teza, Samir Suweis, Marco Gherardi, Amos Maritan, and Marco~Cosentino
  Lagomarsino.
\newblock Network model of conviction-driven social segregation.
\newblock {\em Phys. Rev. E}, 99(3):032310, 2019.

\bibitem{van2005global}
Marshall Van~Alstyne and Erik Brynjolfsson.
\newblock Global village or cyber-balkans? modeling and measuring the
  integration of electronic communities.
\newblock {\em Management Science}, 51(6):851--868, 2005.

\bibitem{vosoughi2018spread}
Soroush Vosoughi, Deb Roy, and Sinan Aral.
\newblock The spread of true and false news online.
\newblock {\em Science}, 359(6380):1146--1151, 2018.

\bibitem{Weng2012}
L~Weng, A~Flammini, Alessandro Vespignani, and Filippo Menczer.
\newblock {Competition among memes in a world with limited attention.}
\newblock {\em Sci. Rep.}, 2:335, 2012.

\bibitem{weng2013role}
Lilian Weng, Jacob Ratkiewicz, Nicola Perra, Bruno Gon\c{c}alves, Carlos
  Castillo, Francesco Bonchi, Rossano Schifanella, Filippo Menczer, and
  Alessandro Flammini.
\newblock The role of information diffusion in the evolution of social
  networks.
\newblock In {\em Proceedings of the 19th ACM SIGKDD International Conference
  on Knowledge Discovery and Data Mining}, KDD '13, pages 356--364, New York,
  NY, USA, 2013. ACM.

\bibitem{windrum2007}
Paul Windrum, Giorgio Fagiolo, and Alessio Moneta.
\newblock Empirical validation of agent-based models: Alternatives and
  prospects.
\newblock {\em Journal of Artificial Societies and Social Simulation}, 10(2):8,
  2007.

\bibitem{yu2017opinion}
Y.~Yu, G.~Xiao, G.~Li, W.~P. Tay, and H.~F. Teoh.
\newblock Opinion diversity and community formation in adaptive networks.
\newblock {\em Chaos: An Interdisciplinary Journal of Nonlinear Science},
  27(10):103115, 2017.

\end{thebibliography}
\end{document}